\documentstyle[12pt,amstex]{article}
  \newtheorem{df}{Definition}
  \newtheorem{th}[df]{Theorem}

  \newtheorem{cor}[df]{Corollary}

\begin{document}
 \title{ Some Useful Formulas \\ in \\
         Nonlinear Sigma Models in $(1+2)$-Dimensions \\
         \quad \\
         \it{To the memory of Kiyoshi Hayashi}}
 \author{ 
  Kazuyuki FUJII\thanks{Department of Mathematics, 
  Yokohama City University, 
  Yokohama 236, 
  Japan, \endgraf 
  {\it E-mail address}: fujii{\char'100}yokohama-cu.ac.jp} \ and 
  Tatsuo SUZUKI\thanks{Department of Mathematics, 
  Waseda University, 
  Tokyo 169, 
  Japan, \endgraf 
  {\it E-mail address}: 695m5050{\char'100}mn.waseda.ac.jp}}
 \date{}
 \maketitle
\begin{abstract}
We give in this paper some formulas which are useful in the construction of nontrivial conserved currents for submodels of $CP^1$-model or $QP^1$-model in $(1+2)$ dimensions. 

These are full generalization of our results in the previous paper (hep-th 9802105).
\end{abstract}
In the previous paper \cite{F-S}, we have given explicit formulas to the nontrivial conserved currents in the submodel of $CP^1$-model \cite{A-F-G} and of other models in $(1+2)$ dimensions.

But the looks of our formulas in \cite{F-S} (Proposition 2.1 and Proposition 4.1) are not so good from the viewpoint of symmetry.

After some work, we could overcome this point. As a by-product, we could generalize our formulas. In this letter, we will report our generalized formulas.

Let $\frak g$ be $\frak s\frak l(\mbox{2,\bf C})$, the Lie algebra of $SL(2,\bf C)$ and $\{ T_{+}, T_{-}, T_3 \} $ be its generators satisfying 
\begin{equation}
 [T_3,T_{+}]=T_{+}, \quad [T_3,T_{-}]=-T_{-}, \quad [T_{+},T_{-}]=2T_3.
\end{equation}
Usually we choose
\begin{equation}
 T_{+}= \left(
          \begin{array}{cc}
            0 & 1 \\
            0 & 0 
          \end{array}
        \right), \quad 
 T_{-}= \left(
          \begin{array}{cc}
            0 & 0 \\
            1 & 0 
          \end{array}
        \right), \quad 
 T_3  = \frac12
        \left(
          \begin{array}{cc}
            1 &  0 \\
            0 & -1 
          \end{array}
        \right) .
\end{equation}
From here, we consider a spin $j$ representation of $\frak g$. Starting from this $\frak g$, we can construct a non-semisimple Lie algebra $\hat{\frak g}$ as follows (see \cite{A-F-G} or \cite{F-S}).
\begin{equation}
 0 \rightarrow P^{(j)} \rightarrow \hat{\frak g} \rightarrow \frak g \rightarrow\mbox{0},
\end{equation}
where $P^{(j)}$ is a representation space of $\frak g$ (abelian ideal of $\hat{\frak g}$). The algebra in $\hat{\frak g}$ is given by
\begin{eqnarray}
 && [T_3, P_m^{(j)}] = mP_m^{(j)}, \nonumber\\
 && [T_{\pm}, P_m^{(j)}] = \sqrt{(j \mp m)(j \pm m+1)}P_{m \pm 1}^{(j)}, \\
 && [P_m^{(j)}, P_n^{(j)}] = 0 \nonumber
\end{eqnarray}
where $m \in \{ -j,-j+1, \cdots, j-1,j \} $ and $\{ P_m^{(j)}|-j \le m \le j \}$ is a set of generators of $P^{(j)} (\cong {\bf C}^{2j+1})$. We note that $P_j^{(j)}$ $(P_{-j}^{(j)})$ is the highest (lowest) spin state.

For $u:M^{1+2} \rightarrow \bf C$, we set
\begin{equation}
 W \equiv W(u)=
          \frac{1}{\sqrt{1+|u|^2}}
           \left(
            \begin{array}{cc}
               1     & iu \\
            i\bar{u} & 1 \\
            \end{array}
           \right)
 \label{eqn:5}
\end{equation}
and choose a gauge field $A_{\mu}$ and an anti-symmetric tensor field $B_{\mu \nu}$ or its dual field $\tilde{B}_{\mu}$ $(=\frac12 \epsilon_{\mu \nu \lambda} B^{\nu \lambda})$ as
\begin{eqnarray}
 A_{\mu} & \equiv & -\partial_{\mu}W W^{-1} \nonumber\\
         & = & \frac{-1}{1+|u|^2}
               \{ i\partial_{\mu}u T_{+}+i\partial_{\mu}\bar{u} T_{-}+
                  (\partial_{\mu}u \bar{u}-u\partial_{\mu}\bar{u})T_3 \}, 
 \label{eqn:6}\\
 \tilde{B}_{\mu}^{(j)} & \equiv & \frac{1}{1+|u|^2}
             (\partial_{\mu}u P_1^{(j)}-\partial_{\mu}\bar{u} P_{-1}^{(j)}).
 \label{eqn:7}
\end{eqnarray}
In the choice of (\ref{eqn:7}), $P_1^{(j)}$ $(P_{-1}^{(j)})$ are no longer the highest (lowest) spin state for $j \geq 2$.

A comment is in order. The Gauss decomposition of $W$ in (\ref{eqn:5}) is given by
\begin{equation}
 W=W_1 \equiv e^{iuT_{+}}e^{\varphi T_3}e^{i\bar{u}T_{-}}
\end{equation}
or
\begin{equation}
 W=W_2 \equiv e^{i\bar{u}T_{-}}e^{-\varphi T_3}e^{iuT_{+}}
\end{equation}
where $\varphi = \log{(1+|u|^2)}$. These expressions will become useful in later calculations. Now, if we assume
\begin{equation}
 D_{\mu}\tilde{B}^{\mu (j)}=0
\end{equation}
(this condition determines a submodel of $CP^1$-model) where $
 D_{\mu} \equiv
  \partial_{\mu}+[A_{\mu}, \quad], $ then
\begin{equation}
 J_{\mu}^{(j)} \equiv W^{-1} \tilde{B}_{\mu}^{(j)} W
   =\sum_{m=-j}^{j} J_{\mu}^{(j,m)}P_m^{(j)} 
\end{equation}
is the conserved currents which we are looking for. Therefore we must determine $\{ J_{\mu}^{(j,m)} | \ |m| \le j \}$ for each $j \geq 1$ \cite{A-F-G}. In fact, in \cite{F-S} we have determined them completely. 

From the form in (\ref{eqn:7}), the most general one of $\tilde{B}_{\mu}^{(j)}$ is 
\begin{equation}
 \tilde{B}_{\mu}^{(j;m)} = \frac{1}{1+|u|^2}
             (\partial_{\mu}u P_m^{(j)}-\partial_{\mu}\bar{u} P_{-m}^{(j)})
 \label{eqn:12}
\end{equation}
where $m \in \{ 1, \cdots ,j \} $. If $m=1$, then $\tilde{B}_{\mu}^{(j;1)}$ reduces to $\tilde{B}_{\mu}^{(j)}$ in (\ref{eqn:7}) and $m=j$ $\tilde{B}_{\mu}^{(j;j)}$ to $\tilde{B}_{\mu}$ in (54) in \cite{F-S}. Now we want to calculate 
\begin{equation}
 J_{\mu}^{(j;m)} \equiv W^{-1} \tilde{B}_{\mu}^{(j;m)} W
   =\sum_{k=-j}^{j} J_{\mu}^{(j;m)}(k)P_k^{(j)}. 
\end{equation}
This calculation is not so easy, but we can perform. Namely
\begin{th}\label{th:1} we have \\
(a) for $k \geq 0$, \\
(i) $0 \leq k \leq m$,
 \begin{eqnarray}
   J_{\mu}^{(j;m)}(k) &=&
    \sqrt{\frac{(j+k)!(j-k)!}{(j+m)!(j-m)!}}
    \frac{1}{(1+|u|^2)^{j+1}} \nonumber\\
  && \hspace{-6mm} \times
   \left\{  
    \sum_{n=0}^{j-m} \alpha _n(m,k) |u|^{2n} (-i\bar{u})^{m-k}\partial_{\mu} u 
   \right. \nonumber\\
  && \hspace{1mm}
   \left.
    -(-1)^{j-m}\sum_{n=0}^{j-m} \alpha _{j-m-n}(m,k) |u|^{2n} (-iu)^{m+k}\partial_{\mu}\bar{u}
   \right\} ,
  \label{eqn:14} 
 \end{eqnarray}
(ii) $m \leq k \leq j$,
 \begin{eqnarray}
   J_{\mu}^{(j;m)}(k) &=&
    \sqrt{\frac{(j+m)!(j-m)!}{(j+k)!(j-k)!}}
    \frac{(-iu)^{k-m}}{(1+|u|^2)^{j+1}} \nonumber\\
  && \hspace{-6mm} \times
   \left\{  
    \sum_{n=0}^{j-k} \alpha _n(k,m) |u|^{2n} \partial_{\mu} u 
   \right. \nonumber\\
  && \hspace{1mm}
   \left.
   - (-1)^{j-k}
    \sum_{n=0}^{j-k} \alpha _{j-k-n}(k,m) |u|^{2n} (-iu)^{2m} \partial_{\mu}\bar{u}
   \right\} ,
  \label{eqn:15} 
 \end{eqnarray}
where 
\begin{equation}
 \alpha _n(m,k) \equiv
   (-1)^n
    \left(
     \begin{array}{c}
      j-m \\
       n  
     \end{array}
    \right)
    \left(
     \begin{array}{c}
       j+m \\
       n+m-k  
     \end{array}
    \right) .
 \label{eqn:16}
\end{equation}
(b) For $k < 0$,
\begin{equation}
  J_{\mu}^{(j;m)}(k)=(-1)^{k+m+1} J_{\mu}^{(j;m)^{\dag}}(-k). 
 \label{eqn:17}
\end{equation}
\end{th}
Let us state some corollaries. First, we set $m=1$ in the theorem.
\begin{cor}\label{cor:2} We have \\
(a) for $k \geq 0$, \\
(i) $k=0$,
 \begin{eqnarray}
   J_{\mu}^{(j;1)}(0) &=&
    \sqrt{\frac{j}{j+1}}
    \frac{-i}{(1+|u|^2)^{j+1}} 
    (\bar{u} \partial_{\mu} u - u \partial_{\mu} \bar{u})
    \nonumber\\
  && \hspace{4mm} \times
    \sum_{n=0}^{j-1} 
    (-1)^n
    \left(
     \begin{array}{c}
      j-1 \\
       n  
     \end{array}
    \right)
    \left(
     \begin{array}{c}
       j+1 \\
       n+1  
     \end{array}
    \right) |u|^{2n} ,
  \label{eqn:18} 
 \end{eqnarray}
(ii) $1 \leq k \leq j$,
 \begin{eqnarray}
   J_{\mu}^{(j;1)}(k) &=&
    \sqrt{\frac{(j+1)!(j-1)!}{(j+k)!(j-k)!}}
    \frac{(-iu)^{k-1}}{(1+|u|^2)^{j+1}} \nonumber\\
  && \hspace{-6mm} \times
   \left\{  
    \sum_{n=0}^{j-k} 
    (-1)^n
    \left(
     \begin{array}{c}
      j-k \\
       n  
     \end{array}
    \right)
    \left(
     \begin{array}{c}
       j+k \\
       n+k-1  
     \end{array}
    \right)
    |u|^{2n} \partial_{\mu} u 
    \right. \nonumber\\
  && \hspace{1mm}
   \left.
    +\sum_{n=0}^{j-k} 
    (-1)^n
    \left(
     \begin{array}{c}
      j-k \\
       n  
     \end{array}
    \right)
    \left(
     \begin{array}{c}
       j+k \\
       n+k+1  
     \end{array}
    \right)
    |u|^{2n} u^2 \partial_{\mu}\bar{u}
   \right\} .
  \label{eqn:19} 
 \end{eqnarray}
(b) For $k < 0$,
\begin{equation}
  J_{\mu}^{(j;1)}(k)=(-1)^k J_{\mu}^{(j;1)^{\dag}}(-k). 
 \label{eqn:20}
\end{equation}
\end{cor}
The appearance between Corollary \ref{cor:2} and Proposition 2.1 in \cite{F-S} seems to be different. But a little algebra shows that
 \begin{eqnarray}
   J_{\mu}^{(j;1)}(0) &=&
    \sqrt{j(j+1)}
    \frac{-i}{(1+|u|^2)^{j+1}} 
    (\bar{u} \partial_{\mu} u - u \partial_{\mu} \bar{u})
    \nonumber\\
  && \hspace{4mm} \times
    \sum_{n=0}^{j-1} 
    (-1)^n
    \frac{1}{j}
    \left(
     \begin{array}{c}
       j \\
       n  
     \end{array}
    \right)
    \left(
     \begin{array}{c}
        j \\
       n+1  
     \end{array}
    \right) |u|^{2n} ,
  \label{eqn:21} 
 \end{eqnarray}
 \begin{eqnarray}
   J_{\mu}^{(j;1)}(k) &=&
    \sqrt{\frac{(j+k)!}{j(j+1)(j-k)!}}
    \frac{(-iu)^{k-1}}{(1+|u|^2)^{j+1}} \nonumber\\
  && \hspace{-14mm} \times
   \left\{  
    \sum_{n=0}^{j-k} 
    (-1)^n
    \frac{n!}{(n+k-1)!}
    \left(
     \begin{array}{c}
      j-k \\
       n  
     \end{array}
    \right)
    \left(
     \begin{array}{c}
       j+1 \\
        n  
     \end{array}
    \right)
    |u|^{2n} \partial_{\mu} u 
    \right. \nonumber\\
  && \hspace{-7mm}
    \left.
    +\sum_{n=0}^{j-k} 
    (-1)^n
    \frac{n!}{(n+k+1)!}
    \left(
     \begin{array}{c}
      j-k \\
       n  
     \end{array}
    \right)
    \left(
     \begin{array}{c}
       j+1 \\
       n+2  
     \end{array}
    \right)
    |u|^{2n} u^2 \partial_{\mu}\bar{u}
   \right\}.
  \label{eqn:22} 
\end{eqnarray}
Namely, they agree if we replace $k$ in (\ref{eqn:22}) with $m$. Comparing (\ref{eqn:18}), (\ref{eqn:19}) with (\ref{eqn:21}), (\ref{eqn:22}), we see that the formulas in Corollary \ref{cor:2} are clearer than those in Proposition 2.1 in \cite{F-S} from the viewpoint of symmetry. 

Next we set $m=j$ in the theorem.
\begin{cor}\label{cor:3} We have \\
(a) for $k \geq 0$, 
 \begin{eqnarray}
   J_{\mu}^{(j;j)}(k) &=&
    \sqrt{\frac{(2j)!}{(j+k)!(j-k)!}}
    \frac{1}{(1+|u|^2)^{j+1}} \nonumber\\
  && \hspace{4mm} \times
   \left\{  
    (-i\bar{u})^{j-k} \partial_{\mu} u 
       -(-iu)^{j+k} \partial_{\mu}\bar{u}
   \right\} .
  \label{eqn:23} 
 \end{eqnarray}
(b) For $k < 0$,
\begin{equation}
  J_{\mu}^{(j;j)}(k)=(-1)^{j+1+k} J_{\mu}^{(j;j)^{\dag}}(-k). 
 \label{eqn:24}
\end{equation}
\end{cor}
This corollary agrees with Proposition 3.1 in \cite{F-S}. Then constants $\alpha , \beta $ in Proposition 3.1 are, for $k > 0$, 
\begin{eqnarray}
 \alpha &=& \sqrt{\frac{(2j)!}{(j+k)!(j-k)!}} (-i)^{j-k}, \\
 \beta &=& -\sqrt{\frac{(2j)!}{(j+k)!(j-k)!}} (-i)^{j+k}.
\end{eqnarray}
A comment is in order. We in (\ref{eqn:16}) defined $\alpha_n$ as
$$
 \alpha_n(m,k) =
   (-1)^n
    \left(
     \begin{array}{c}
      j-m \\
       n  
     \end{array}
    \right)
    \left(
     \begin{array}{c}
       j+m \\
       n+m-k  
     \end{array}
    \right) .
$$
For this, we have simple relations:
\begin{eqnarray}
 \alpha_n(m,-k) &=& \frac{(j+m)!(j-m)!}{(j+k)!(j-k)!} \alpha_n(k,-m),\\
 \alpha_{j-m-n}(m,k) &=& (-1)^{j-m} \alpha_n(m,-k).
\end{eqnarray}
These formulas are also useful in our calculations.

Finally, we consider the case corresponding to $QP^1$-model. 
For $u:M^{1+2} \rightarrow D$, where $D$ is the Poincare disk in {\bf C}, 
we set
\begin{equation}
 W \equiv W(u)=
          \frac{1}{\sqrt{1-|u|^2}}
           \left(
            \begin{array}{cc}
               1     & iu \\
            -i\bar{u} & 1 \\
            \end{array}
           \right) .
\end{equation}
For this, the Gauss decomposition is given by
\begin{equation}
 W=W_1 \equiv e^{iuT_{+}}e^{\varphi T_3}e^{-i\bar{u}T_{-}}
\end{equation}
or
\begin{equation}
 W=W_2 \equiv e^{-i\bar{u}T_{-}}e^{-\varphi T_3}e^{iuT_{+}}
\end{equation}
where $\varphi = \log{(1-|u|^2)}$. As in (\ref{eqn:12}), we set
\begin{equation}
 \tilde{B}_{\mu}^{(j;m)} = \frac{1}{1-|u|^2}
             (\partial_{\mu}u P_m^{(j)}+\partial_{\mu}\bar{u} P_{-m}^{(j)})
 \label{eqn:32}
\end{equation}
for $m \in \{ 1, \cdots ,j \} $. Similarly in the preceding case, what we want to calculate is
\begin{equation}
 J_{\mu}^{(j;m)} \equiv W^{-1} \tilde{B}_{\mu}^{(j;m)} W
   =\sum_{k=-j}^{j} J_{\mu}^{(j;m)}(k)P_k^{(j)}. 
\end{equation}
For this case, we can use Theorem \ref{th:1}. Namely, if we replace
\begin{equation}
 u \rightarrow u, \quad \bar{u} \rightarrow -\bar{u}
\end{equation}
in Theorem \ref{th:1} (of course $|u| < 1$), then
\begin{th}\label{th:4} we have \\
(a) for $k \geq 0$, \\
(i) $0 \leq k \leq m$,
 \begin{eqnarray}
   J_{\mu}^{(j;m)}(k) &=&
    \sqrt{\frac{(j+k)!(j-k)!}{(j+m)!(j-m)!}}
    \frac{1}{(1-|u|^2)^{j+1}} \nonumber\\
  && \hspace{-6mm} \times
   \left\{  
    \sum_{n=0}^{j-m} \tilde{\alpha}_n(m,k) |u|^{2n} (i\bar{u})^{m-k}\partial_{\mu} u 
   \right. \nonumber\\
  && \hspace{1mm}
   \left.
   +\sum_{n=0}^{j-m} \tilde{\alpha}_{j-m-n}(m,k) |u|^{2n} (-iu)^{m+k}\partial_{\mu}\bar{u}
   \right\} ,
  \label{eqn:35} 
 \end{eqnarray}
(ii) $m \leq k \leq j$,
 \begin{eqnarray}
   J_{\mu}^{(j;m)}(k) &=&
    \sqrt{\frac{(j+m)!(j-m)!}{(j+k)!(j-k)!}}
    \frac{(-iu)^{k-m}}{(1-|u|^2)^{j+1}} \nonumber\\
  && \hspace{-6mm} \times
   \left\{  
    \sum_{n=0}^{j-k} \tilde{\alpha}_n(k,m) |u|^{2n} \partial_{\mu} u 
   \right. \nonumber\\
  && \hspace{1mm}
   \left.
   +\sum_{n=0}^{j-k} \tilde{\alpha}_{j-k-n}(k,m) |u|^{2n} (-iu)^{2m} \partial_{\mu}\bar{u}
   \right\} ,
  \label{eqn:36} 
 \end{eqnarray}
where 
\begin{equation}
 \tilde{\alpha}_n(m,k) \equiv
    \left(
     \begin{array}{c}
      j-m \\
       n  
     \end{array}
    \right)
    \left(
     \begin{array}{c}
       j+m \\
       n+m-k  
     \end{array}
    \right) .
 \label{eqn:37}
\end{equation}
(b) For $k < 0$,
\begin{equation}
  J_{\mu}^{(j;m)}(k)=J_{\mu}^{(j;m)^{\dag}}(-k). 
 \label{eqn:38}
\end{equation}
\end{th}
In this theorem, we set $m=1$.
\begin{cor}\label{cor:5} We have \\
(a) for $k \geq 0$, \\
(i) $k=0$,
 \begin{eqnarray}
   J_{\mu}^{(j;1)}(0) &=&
    \sqrt{\frac{j}{j+1}}
    \frac{i}{(1-|u|^2)^{j+1}} 
    (\bar{u} \partial_{\mu} u - u \partial_{\mu} \bar{u})
    \nonumber\\
  && \hspace{4mm} \times
    \sum_{n=0}^{j-1} 
    \left(
     \begin{array}{c}
      j-1 \\
       n  
     \end{array}
    \right)
    \left(
     \begin{array}{c}
       j+1 \\
       n+1  
     \end{array}
    \right) |u|^{2n} ,
  \label{eqn:39} 
 \end{eqnarray}
(ii) $1 \leq k \leq j$,
 \begin{eqnarray}
   J_{\mu}^{(j;1)}(k) &=&
    \sqrt{\frac{(j+1)!(j-1)!}{(j+k)!(j-k)!}}
    \frac{(-iu)^{k-1}}{(1-|u|^2)^{j+1}} \nonumber\\
  && \hspace{-6mm} \times
   \left\{  
    \sum_{n=0}^{j-k} 
    \left(
     \begin{array}{c}
      j-k \\
       n  
     \end{array}
    \right)
    \left(
     \begin{array}{c}
       j+k \\
       n+k-1  
     \end{array}
    \right)
    |u|^{2n} \partial_{\mu} u 
   \right. \nonumber\\
  && \hspace{1mm}
   \left.
   -\sum_{n=0}^{j-k} 
    \left(
     \begin{array}{c}
      j-k \\
       n  
     \end{array}
    \right)
    \left(
     \begin{array}{c}
       j+k \\
       n+k+1  
     \end{array}
    \right)
    |u|^{2n} u^2 \partial_{\mu}\bar{u}
   \right\} .
  \label{eqn:40} 
 \end{eqnarray}
(b) For $k < 0$,
\begin{equation}
  J_{\mu}^{(j;1)}(k)=J_{\mu}^{(j;1)^{\dag}}(-k). 
 \label{eqn:41}
\end{equation}
\end{cor}
Similarly in the preceding case, it is easy to see that Corollary \ref{cor:5} agrees with Proposition 4.1 in \cite{F-S}.
 \section*{Acknowledgements}
We are very grateful to Yoshinori Machida for warm hospitality during our stay in Numazu technical college.

  
\end{document}